# A H-K CLUSTERING ALGORITHM FOR HIGH DIMENSIONAL DATA USING ENSEMBLE LEARNING

Rashmi Paithankar[1] and Bharat Tidke[2]

[1]Department of Computer Engg
Flora Institute of Technology, Pune
Maharashtra, India
[2]Assistant Professor, Department of Computer Engg
Flora Institute of Technology, Pune
Maharashtra, India

## ABSTRACT

*Advances made to the traditional clustering algorithms solves the various problems such as curse of dimensionality and sparsity of data for multiple attributes. The traditional H-K clustering algorithm can solve the randomness and apriority of the initial centers of K-means clustering algorithm. But when we apply it to high dimensional data it causes the dimensional disaster problem due to high computational complexity. All the advanced clustering algorithms like subspace and ensemble clustering algorithms improve the performance for clustering high dimension dataset from different aspects in different extent. Still these algorithms will improve the performance form a single perspective. The objective of the proposed model is to improve the performance of traditional H-K clustering and overcome the limitations such as high computational complexity and poor accuracy for high dimensional data by combining the three different approaches of clustering algorithm as subspace clustering algorithm and ensemble clustering algorithm with H-K clustering algorithm.*

## 1. INTRODUCTION

As an important technique in data mining, clustering analysis groups the observations having similar properties which can be called as an unsupervised classification[1] which helps to extract the relevant information from high dimensional data. Hierarchical clustering and partition clustering are the basic types of clustering algorithms. Hierarchical clustering seeks to build a hierarchy of clusters which can be formed by using single link and complete link clustering algorithms. It does not require to pre specify the number of clusters. Examples for these algorithms are BRICH (Balance Iterative Reducing and Clustering using Hierarchies) and CURE (Cluster Using Representatives). Another important type of clustering is Partition clustering, which obtains a single partition of the data instead of clustering structure. It uses criteria function optimization to create clusters locally or globally [1]. Partition cluster have advantage in large applications but we have to pre specify the number of desired output clusters. The K-means algorithm is the most typical partition algorithm, which is quite popular as it is easy to implement and does not require user to specify many parameters.

Applying the traditional clustering algorithms on the high dimensional datasets regularly presented a great challenge for traditional data mining techniques both in terms of effectiveness and efficiency. Increasing sparsity of data and increasing difficulty in distinguishing distances between data points which is due to the so called 'dimensionality disaster' makes clustering difficult. So, adaptations to existing algorithms are required to maintain data quality and speed. Research in the area of clustering introduced a lot of new concepts as subspace clustering, ensemble clustering, and H-K clustering algorithm. The traditional H-K clustering algorithm can solve the randomness and apriority of the initial centers of K-means clustering algorithm. However, it will lead to a dimensional disaster problem when apply to high dimensional dataset clustering due to its high computational complexity and provides clusters with poor accuracy. Subspace clustering which is an extension of traditional clustering, finds the clusters in various datasets [8] and provides scalability, end user comprehensibility of the results, non-presumption, insensitivity to the order of input records, accuracy and speed also removes redundancy and find overlapping clusters in the subspaces[4,5,6,7,9,10]. Ensemble clustering 'the knowledge reuse framework', firstly proposed by Strel and Ghosh [11] is the technique which uses the two mechanisms as generation mechanism which generates the clusters using different criteria and consensus function will choose the most appropriate solution form the set of solutions . It overcome the challenges created by high dimensional data and gives high performance on real world datasets for applications as Internet applications and medical diagnostics [2,3,12,13,19,20]. The proposed model combines the three techniques, subspace clustering, H-K clustering and ensemble clustering and their advantages to improve the performance of clustering result on high dimensional data which will simultaneously overcome the limitations of H-K clustering algorithm for high dimensional data ( as high computational complexity and poor accuracy).

## 2. MOTIVATION

The traditional algorithms for clustering does not give the effective and efficient results when we want to deal with high dimensional data as it has the disadvantages such as the "curse of dimensionality" and the "empty space phenomenon". In high dimensional spaces, the data are inherently sparse, and the distance between each pair of points is almost the same for a wide variety of data distributions and distance functions [4]. Meanwhile, the notion of density is even more troublesome than that of distance. These problems can be referred to as the "curse of dimensionality". To overcome these problems of irrelevant and noisy features and sparsity of data it is important to provide advanced clustering algorithm that will solve the above problems and cluster the data efficiently. Proposed model has provided with the combination of advanced clustering algorithms that will improve the cluster quality and speed.

## 3. RELATED WORK

A lot of work has been done in the area of clustering, based on the research until date, the general categorization for high dimensional data set clustering includes: 1- Dimension reduction, 2- Subspace clustering, 3 - Ensemble Clustering and 4 - H-K clustering [1] [11] [14]. Following section gives an overview and some of the limitations of the above techniques.

### 3.1. Dimension reduction

Feature selection and feature transformation are the most popular techniques of dimension reduction [5]. Feature transformation techniques create a combination of multiple attributes and make summary of them. [5]. These methods include techniques such as principle component analysis and singular value decomposition. Feature selection methods reveal groups of objects having the similar attributes by picking up the most relevant attributes form the dataset [5]. Yanchang et al. [16] proposed a method in which he used transformation technique and break the high dimensional clustering into several one or two dimensional clustering phases and apply common clustering algorithms on them. Experiments with different datasets showed that, the time complexity of clustering can be linear with the dimensionality of datasets. This framework can easily process hybrid datasets but may face problems for datasets containing overlapping clusters. Chen et al. [17] proposed IMSND (Initialization Method based on Shared Neighborhood Density) which is a local density based method used to find the probability density of a point to search for initial cluster centers on high dimensional data. Author implemented this method on the spherical K-Means algorithm. An experimental evaluation shows the increased performance of K-Means algorithm. But in both methods (Feature Selection and transformation) we will have losing information which naturally affects accuracy [16, 17] and feature selection algorithms have difficulty when clusters are found in different subspaces. This type of data motivated the evolution of subspace clustering algorithms.

### 3.2. Subspace clustering

Subspace clustering is an extension of traditional clustering which finds the clusters that exist in multiple or possibly overlapping clusters [8]. Bottom up approach and Top down approach are two major kinds of subspace clustering based on search strategy. Top down algorithms makes the use of full set of dimensions and reveal the set of subspaces iteratively which starts from an initial set of subspace [8]. Example algorithms are CLIQUE, ENCLUS, and MAFIA etc. Bottom up approaches consider each object as a separate cluster and combine them to form clusters [8]. Example algorithms are PROCLUS, ORCLUS, and PREDECON etc.

Agrawal et al. [10] proposed a clustering algorithm 'CLIQUE' which identifies dense clusters in subspaces of maximum dimensionality that satisfies the requirements of data for data mining applications as scalability, end user comprehensibility of the results, non-presumption, and insensitivity to the order of input records but does not evaluate the quality of clustering in different subspaces. Chen et al. [6] presented a technique for solving the problem of selecting the k representative clusters by examining the relationship between low dimensional subspace clusters and high dimensional ones by using an approximate method 'PCoC'. Muller et al. [12] presented a novel model called 'RESCU', which extracts the most interesting, non - redundant clusters by using global optimization and provide a proof that proved this problem as NP- hard. Kriegel et al. [7] Proposed finding overlapping clusters in the subspaces, by using the filter refinement architecture, which speed up the subspace finding process and scales at most quadratic w. r. t. to the data dimensionality and subspace dimensionality. Proposed approach overcomes the problems of exponentially scaling of algorithms with the data or subspace dimensionality and the problems caused by the use of global density threshold for clustering. Input data is preprocessed by using the algorithms as DBSCAN, K-Means, and SNN, which finds the base clusters. After words base clusters are merged to find maximal dimensional cluster approximation on which post processing (Pruning, Refinement etc) is applied. Ali et al. [5] proposed a method based on divide and conquers technique, which is a two step clustering. First it select the subspaces based on size/level and again perform clustering on that subspaces based on similarity that uses K-means algorithm. This method improves accuracy and efficiency of original K-means algorithm.

### 3.3. Ensemble Clustering

Ensemble clustering combines the solutions of multiple clustering algorithms using the consensus function and form the more relevant solution. General procedure for it is shown in fig. 1

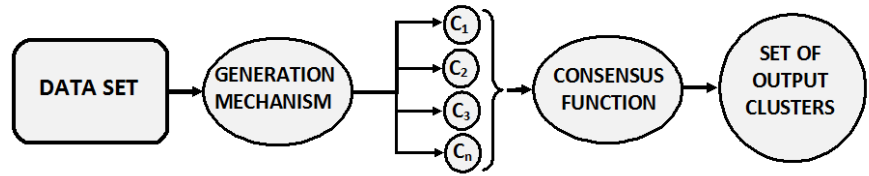

Fig. 1 Ensemble clustering process

Tidke et al. [3] proposed a clustering ensemble method based on two staged clustering algorithm to overcome the challenges created by high dimensional data such as, curse of dimensionality and the problem of visualizing high dimensional data in certain cases. PROCLUS is used for initial subspace clustering, K-means partitioning algorithm is applied on generated subspaces which is followed by split and merge techniques in which threshold value, distance function and mean square error conditions are considered respectively. Zhizhou KONG et al. [12], proposed the mechanism of the integrate mechanism of testing and information rolling method to decrease the error probability of matching cluster members, sand using a method of category weight to ensemble clustering. Clusters members are generated using Ward's Method, k-means Method and Median Method. P. Viswanath et al. [13] presented an ensemble of leaders clustering methods where the entire ensemble requires only a single scan of the data set. 'Deferring buffer scheme' which is an improvement over 'Blocked access scheme' used for accessing data from dataset and Consensus function called 'Co association Matrix' is used to ensemble individual partitions. Weiwei Zhuang [19], Derek Greene [20] applied the ensemble clustering algorithm for the real time applications data such as Internet applications and medical diagnostics respectively.

### 3.4. H-K clustering

H-K clustering algorithm is proposed and implemented for deciding the k clusters for k-means algorithm. It is implemented in divisive H-K and agglomerative H-K clustering. Divisive H-K algorithm implements a top-down approach which splits the whole dataset into the small clusters. It divides the K clusters into K+1 clusters using K-means method. Agglomerative clustering works by merging the small clusters together. It merges the K clusters into K-I clusters.

In 2005, Tung-Shou Chen et al. [15] proposed H-K (Hierarchical K-means clustering algorithm) clustering algorithm, combining hierarchical clustering method and partition clustering method organically for data clustering. Compared with single algorithm, H-K clustering algorithm can solve the problem of randomness and apriority of initial centers selection in k-means clustering process, and obtain better clustering result. But it is a pity that it still needs high computing complexity.

Ying HE et al. [14] proposed ensemble learning for high dimension data clustering, and proposes a new clustering algorithm named EPCAHK clustering algorithm(Ensemble Principle Analysis Hierarchical K-means clustering algorithm, EPCAHK), which helps to improve the performance of traditional H-K clustering algorithm in high dimensional datasets. Firstly the high dimensional dataset is converted to low dimensional using PCA data reduction technique. Subsequently, the clustering results of the hierarchical stage for obtaining initial information (e.g., the cluster number or the initial clustering centers) are integrated by using the min-transitive closure method. Finally, the final clustering result is achieved by using K-means clustering algorithm based on the ensemble clustering results, and provides some issues which need to be addressed in the future as the relationship between ensemble size and the ensemble clustering algorithm performance, distribution of the dataset and the clustering performance.

Table I comparative analysis of techniques for high dimensional data clustering

| Author | Clustering Technique | Method | Observation |
|---|---|---|---|
| Yanchang et al.[16] | Dimension Reduction | - Convert high dimensional data to Low dimension<br>- Common clustering algorithms | - Improves performance<br>- Loses information<br>- Difficult to find clusters in different subspaces |
| Chen et al.[17] | Dimension Reduction | - IMSND<br>- Spherical K-Means algorithm | |
| Agrawal et al.[10] | Subspace clustering | CLIQUE - identifies dense clusters in subspaces of maximum dimensionality | - Provides scalability, end user comprehensibility of the results, non-presumption, insensitivity to the order of input records<br>- Improves accuracy and speed |
| Chen et al.[6] | Subspace clustering | - A technique for solving the problem of selecting the k representative clusters | |
| Muller et al.[12] | Subspace clustering | - Extracts the most interesting, non - redundant clusters | -Removes redundancy |
| Kriegel et al. [7] | Subspace clustering | - Filter refinement architecture | -Find overlapping clusters in the subspaces<br>- Speed up the subspace finding process |
| Tidke et al.[3] | Ensemble clustering | -Two staged clustering algorithm<br>- PROCLUS | -Overcome the challenges created by high dimensional data |
| Zhizhou KONG et al. [12] | Ensemble clustering | - Category weight method | -Decrease the error probability of matching cluster members |
| Weiwei Zhuang [19] Derek Greene [20] | Ensemble clustering | Apply on real time applications data such as Internet applications and medical diagnostics | |
| Tung-Shou Chen et al.[15] | H-K clustering | -Combine hierarchical clustering method and partition clustering method | -Removes randomness and apriority of initial centers selection in k-means clustering process<br>- High computing complexity |
| Ying HE et al. [14] | H-K clustering | -Ensemble Principle Analysis Hierarchical K-means clustering algorithm- EPCAHK | -Improves clustering performance |

## 4.PROPOSED MODEL

The flow chart of proposed model is shown in the below,

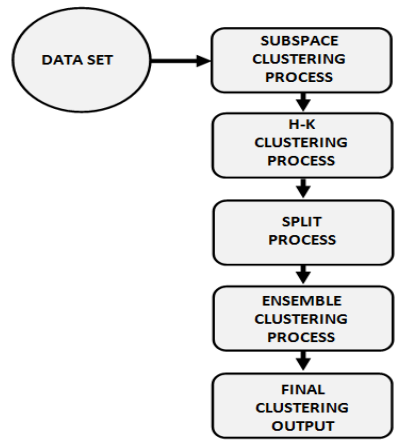

Fig. 2 Block diagram of H-K clustering algorithm based on ensemble learning

Based on the above flowchart of proposed model, the following content will unfold these five sub-stages in details:

**Stage1. Dataset preprocessing**

Import the dataset for clustering and get the information of it as, number of clusters k and the sample number N and the threshold value. And if the dataset contain any missing values replace it with zero and obtain the pre-processed dataset D. (Mostly prefer the real time datasets for more accurate results)

**Stage2. The subspace clustering process**

Adopt the subspace clustering algorithm - ORCLUS, on the pre-processed dataset D which will take the dataset through following steps and the output will give the subsets of dataset. The ORCLUS algorithm is mainly divided into three steps: assign clusters, subspace determination and merge. Assign phase select the k centers and assign the points iteratively to the nearest centers, which is followed by the subspace determination phase that will find the subspace $E_i$ of dimensionality d by calculating the covariance matrix for cluster $C_i$ and selecting the d orthogonal eigen vectors having the least eigen value ( i.e. least spread). Finally the merge stage reduces the number of clusters by combining the clusters which are closer and similar to each other and having least spread.

**Stage3. The H-K clustering process**

Adopt the H-K clustering on the k subspaces which is the output of stage 2. Apply the divisive HK means clustering, on the k subspaces it divides the k cluster dataset into k+1 clusters using k means method. This will help pick up the two elements that are furthest from each other in this cluster, so as to divide the distance between the two into 3 equivalent parts to produce one more new cluster. Repeat this process for the range of [2, k+10] and by applying the random selection method select the L cluster as the output and store them as H(1),H(2),H(3). . . . . , H(L).

**Stage4. Split clustering process**

This stage follows a split strategy based on a predefined threshold value. If the size of the $l^{th}$ cluster is greater than the threshold then split the cluster into two new clusters depending upon the distance between each point and the two centroids of the two new clusters (Centroid is the mean of each cluster). The distance between point and cluster is calculated using Euclidean distance. Again repeat the above procedure of the newly created cluster, run out of the predefined threshold value. Above steps are given below in the form of pseudo code.

**Algorithm 1**:

Input: k clusters and threshold T
Output: n cluster
1. Start with k clusters.
2. Check density of each cluster for given threshold T.
3. If density is more than threshold split the cluster into two based on the distance each point is assign to its closest centroid.

$$J = \sum_{j=1}^{k}\sum_{i-1}^{x}||xi(j) - cj||2 \quad (1)$$

Where $|| x_i^{(j)} - c_j ||^2$ is a chosen distance measure between a data point $x_i^{(j)}$ and $c_j$ the cluster centre, is an indicator of the distance of the *n* data points from their respective cluster centers.
4. Repeat it for each cluster till it reaches threshold value. Now when hierarchy of cluster with similar in size formed by splitting phase merging is required to find out the closest cluster to be merge.

**Stage5. Ensemble clustering process**

After splitting the cluster step cluster adopt the merging and merge the closest cluster based on an objective function. Proposed model uses distance function as an objective function to merge the nearly cluster. In proposed method, child cluster from any parent cluster can be merged, if there distance is smaller than other cluster in the hierarchy. Also check mean square error(MSE) of each merged cluster with the parent cluster if found to be larger, that cluster must be unmerged and available to be merge with some other cluster in the hierarchy this process repeats until all MSE of all possible combination of merged cluster is checked with its parent cluster. Finally the number of cluster merged and remain are the output cluster. The algorithm steps are given below:

**Algorithm 2:**

Input: hierarchy of cluster
Output: partition C1….Cn
1. Start with n node cluster.
2. Find the closest two cluster using Euclidean distance from the hierarchy and merge them
3. Calculate MSE of root cluster and new merge cluster

$$SSE = \sum_{j=1}^{k}\sum_{xi \in cj}||xi - \mu j||2xi - \mu j|| \quad (2)$$

Where, $\mu j$ is the mean of cluster Cj and x is the data object belongs to Cj cluster. Formula to compute $\mu j$ is shown in equation (3).

$$\mu j = (1/nj)\sum xi \in cj\, xi \quad (3)$$

In sum of squared error formula, the distance from the data object to its cluster centroid is squared and distances are minimized for each data object. Main objective of this formula is to generate compact and separate clusters as possible
4. If MSE of new merge cluster is smaller than the cluster after splitting keep it otherwise unmerges them.
5. Repeat until all possible clusters are merged according to step 4.

The above steps can be applied on the high dimensional datasets such as,

1.Cancer (Breast) dataset having 9 attributes as Age,Menopause,Tumorsize,Invnodes,Nodecaps,Degmalig,Breast,Breast-quad,Irradiat and
2.Wdbc dataset having 9 attributes as Clump thickness, Uniformity of cell size, Uniformity of cell shape, Amount of marginal adhesion, Frequency of bare nuclei, Single epithelial cell size, Bland chromatin, Normal nucleoli, Mitoses.

## 5.CONCLUSION

High dimensional dataset processing faces many problems such as 'curse of dimensionality' and 'the sparsity of data in the high dimensional space'. The proposed model provides a solution algorithm for processing the high dimensional dataset which is a combination of the three approaches and makes use of the advantages of ensemble and subspace clustering and simultaneously overcomes the limitations of the traditional H-K clustering such as, high computational complexity and poor accuracy by providing a three stage clustering process in which, firstly the dataset D is converted to subspaces using the subspace clustering algorithm (ORCLUS). Each subspace in the output will reveal the different characteristics of the original dataset. Considering each subspace as the different dataset, adopt the hierarchical clustering. Subsequently, the clustering results of the hierarchical stage is again passed to the split stage in which clusters that are above the threshold of size are split into new clusters, and finally they are integrated by using the objective function (MSE). Applying the various clustering approaches simultaneously will help to improve the performance of clustering process and will provide the stability of H-K clustering algorithm for high dimensional data.